\documentstyle[12pt,epsf]{article}
\input epsf.tex
\newcommand{\be}{\begin{equation}}
\newcommand{\ee}{\end{equation}}
\newcommand{\bea}{\begin{eqnarray}}
\newcommand{\eea}{\end{eqnarray}}
\newcommand{\nn}{\nonumber}
\newcommand{\vx}{{\bf x}}
\newcommand{\vR}{{\bf R}}
\newcommand{\vr}{{\bf r}}
\newcommand{\vt}{{\bf \Theta}}
\newcommand{\vk}{{\bf K}}
\newcommand{\vp}{{\bf p}}
\newcommand{\vs}{{\bf \sigma}}
\newcommand{\om}{\omega}
\begin{document}
\title{Three Body Bound State in Non-Commutative Space}
\author{M. Haghighat\thanks{email: mansour@cc.iut.ac.ir}\ \ \
and \ \ F. Loran\thanks{email:
loran@cc.iut.ac.ir}\\ \\
  {\it Department of  Physics, Isfahan University of Technology (IUT)}\\
{\it Isfahan,  Iran,} \\{\it and}\\
  {\it Institute for Studies in Theoretical Physics and Mathematics (IPM)}\\
{\it P. O. Box: 19395-5531, Tehran, Iran.}}
\date{}
\maketitle

\begin{abstract}
The Bethe-Salpeter equation in non-commutative QED (NCQED) is
considered for three-body bound state. We study the
non-relativistic limit of this equation in the instantaneous
approximation and derive the corresponding Schr\"{o}dinger
equation in non-commutative space. It is shown that the
experimental data for Helium atom puts an upper bound on the
magnitude of the parameter of non-commutativity,
$\theta\sim10^{-9}\lambda_e^2$.
\end{abstract}
\newpage
Non-commutative spaces \cite{Witten} and their phenomenological
aspects [2-10] have been recently considered by many authors.
Among the others P. M. Ho and H. C. Kao in their paper \cite{Ho}
has claimed that there is not any non-commutative correction to
the Hydrogen atom spectrum at the tree level owing to the
opposite non-commutativity of the particles with opposite
charges. Following these arguments, to detect the effect of
non-commutativity at the tree level one should consider the bound
states including three or more particles. A question here is the
accuracy of experimental data. Fortunately atomic transitions
$2^3P_1\to 2^3P_0$ and $2^3P_1\to 2^3P_2$ in Helium atom  is
recently measured \cite{Storry} (and theoretically calculated
\cite{Zhang}) with an accuracy about 1 kHz as precise as
hyper-fine splitting in positronium. Due to long life time of
these transitions, $98 ns$, Helium atom is considered as one of
the best labs to test QED. Therefore, it seems reasonable to study
the effect of non-commutativity on the spectrum of Helium atom.
For this purpose we calculate the corresponding corrections on
the spectrum of the three body bound state up to the order
$\theta\alpha^4$ by using the Bethe-Salpeter equation (BS) in the
framework of NCQED.
\par
\begin{figure}
\centerline{\epsfxsize=5in\epsffile{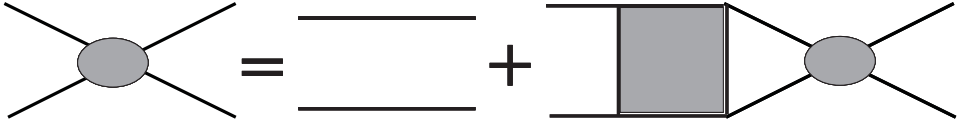}}
\caption{Bethe-Salpeter equation for two particles bound state.}
\label{BS1}
\end{figure}
\begin{figure}
\centerline{\epsfxsize=5in\epsffile{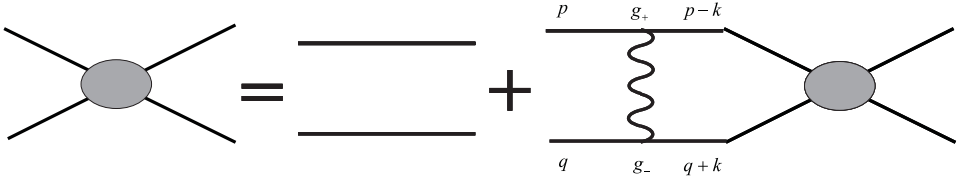}}
\caption{Bethe-Salpeter equation in the ladder approximation. The
coupling constant $g_i= iq_i\gamma_\mu e^{-\frac{i}{2} {\bf
p}.\theta^i.({\bf p}-{\bf k})}$, where $\theta^i$, $i=1,2$ are
the non-commutativity parameters corresponding to the particles
with charges $q_i$.} \label{BS2}
\end{figure}

First we extend the  non-relativistic results of reference
\cite{Ho}, regarding the  lack of  corrections to the spectrum of
Hydrogen on noncommutative spaces, to the full non-relativistic
regime using the BS equation \cite{sal}. The BS equation for the
bound state of a fermion-(anti)fermion system has the form (Fig.
\ref{BS1}):
 \be
 \Psi(p;P)=S(\frac{P}{2}+p)S(\frac{P}{2}-p)\int \frac{d^4k}{(2\pi)^4i} I(k,p) \Psi(p+k;P),
 \label{p1}
 \ee
 where $P$ is the momentum of the centre of mass,
 $\Psi(p;P)$ is the BS amplitude for the bound state and $S(p)$
 is the fermion field propagator. $I(k,p)$ is
 the kernel of the interaction which is the sum of all possible
 irreducible graphs. The effects of non-commutativity are
introduced in QED by modifying the vertices and including
3-photon and 4-photon interactions. Therefore, in non-commutative
QED (NCQED) the kernel of Eq.(\ref{p1}) can be easily constructed
by using the Feynman rules of the theory which are completely
given in references \cite{Ria,Hay}. For instance, for the
Hydrogen atom the tree level calculations correspond to the
ladder approximation for the kernel with free particle
propagators which is given in Fig. \ref{BS2}. At this level one
has
 \be
 I^{l}_{\theta}(k,p)=e^{\frac{i}{2}
 p.(\theta^1+\theta^2).k}I^{l}(k,p),
 \label{p2}
 \ee
 where the superscript $l$ stands for ladder and $I^{l}(k,p)$ is the
interaction kernel in the commutative space in the ladder
approximation. The parameter of non-commutativity $\theta$, is an
antisymmetric real valued tensor defined as follows:
 \be
 \theta^{\mu\nu}=-i\left[x^\mu,x^\nu\right],
 \ee
 Of course $\theta^{0i}$ is assumed to be vanishing since
 $\theta^{0i}\ne 0$ leads to some problems with the concepts of
 causality and the unitarity of field theories \cite{seiberg}.
 The superscripts introduced in Eq.(\ref{p2}) for $\theta$ refer to
 the possibility that in an effective NCQED that considers the
 interactions between particles with internal structures,
 different non-commutativity parameters should be
 assigned to different particles. Furthermore, one can easily see that when
 $\theta_1=-\theta_2$, there is not any non-commutative effect in
 tree level (see Eq.(\ref{p2})).

\par
 In the case of positronium,
 there is also an annihilation diagram in the lowest order. In
such a diagram, each vertex contains a factor $e^{-\frac{i}{2}{\bf
p}.\theta.(-{\bf p})}=1$, thus there is not any non-commutativity
correction at this level due to the annihilation graph.
\par
The BS formalism
 for three body bound state is very complicated. But here, since
 we are interested in lowest order diagrams, the kernel of the BS equation reduces to
 three copies of two body kernel. At the tree level, the kernel
 is,
 \be
 I\Psi=(I_{ope}+I_{ann})\Psi,
 \label{pp2}
 \ee
 where $ope$ stands for one photon exchange and $ann$ for one photon
annihilation interaction. Of course, there is no annihilation
diagram in the case of three body bound states like the Helium
atom. The $I_{ope}$ contains three terms, shown in Fig. 3, as
follows:
 \be
 I_{ope}=I_1+I_2+I_3,
 \label{new1}
 \ee
 \begin{figure}
 \centerline{\epsfxsize=4in\epsffile{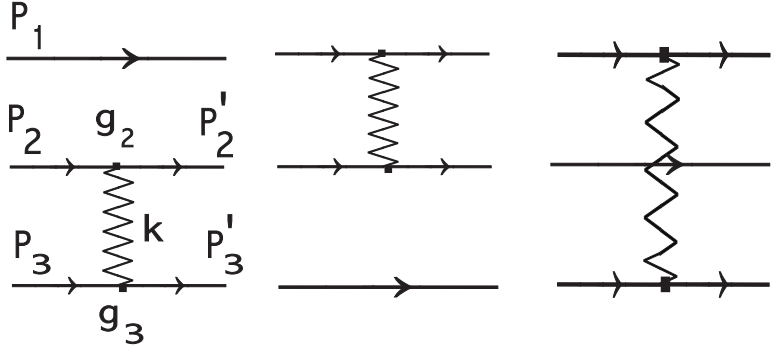}} \caption{One photon exchange diagrams.
 The coupling
 constant $g_i=iq_i\gamma_\mu e^{-\frac{i}{2}{\bf p}_i.\theta^i.{\bf
 p}'_i}$, where $\theta^i$'s are the
non-commutativity parameters corresponding to the particles with charge
$q_i$.}\label{BS3}
 \end{figure}
 where for example
 \be
 I_1=q_2q_3\left(\frac{\gamma^0_2\gamma^0_3}{{{\bf K}}^2}
 -(\frac{\gamma^0_2\gamma^0_3{k_0}^2}{k^2{{\bf K}}^2}-
 \frac{{\bf\gamma}_2\cdot{\bf \gamma}_3}{k^2})\right).
 \label{p0}
 \ee
 To obtain the above relation, we have used Feynman gauge and
 separated the instantaneous interaction from retarded one.
 In the present work, we do not consider the retarded interaction and
 the annihilation diagram since they result
 in higher order corrections in comparison with the instantaneous interaction.
  It is straightforward to show that considering only the instantaneous interaction, in
 non-relativistic approximation, the BS equation with the
 kernel (\ref{new1}), leads to the Schr\"{o}dinger equation for three body bound state.
 To this end we consider the BS equation as follows:
 \be
 \Psi(p_1,p_2;P)=S_1S_2S_3\int \frac{d^4p'_1}{(2\pi)^4i} \frac{d^4p'_2}{(2\pi)^4i}
 I(p_1,p_2;p'_1,p'_2;P)\Psi(p'_1,p'_2;P),
 \label{form1}
  \ee
 where $P$ is the momentum of the centre of mass. The propagators at lowest order are,
 \be
 S_i=\frac{2m_i}{p_i^2-m_i^2+i\epsilon}\hspace{1cm}i=1,2,3,
 \label{form2}
 \ee
 in which $p_3=P-p_1-p_2$. The instantaneous interaction is
 \be
 I_{inst}=\sum_{i<j}\frac{q_iq_j}{\left|{\bf K}_{ij}\right|^2},
 \ee
 where $q_i$, $i=1,2,3$, stand for the charges of the particles and
 ${\bf K}_{ij}$ is the three-momentum of the exchanged photon
 between the $i$th and $j$th lines. Defining
 \be
 \phi(\vp_1,\vp_2)\equiv\int dp^0_1dp^0_2\Psi(p_1,p_2;P),
 \label{form3}
 \ee
 Eq.(\ref{form1}) can be written as follows,
 \be
 \phi(\vp_1,\vp_2)=Z(\vp_1,\vp_2)(V\phi)(\vp_1,\vp_2),
 \label{form4}
 \ee
 where
 \be
 Z(\vp_1,\vp_2)=\int dp^0_1dp^0_2 S_1S_2S_3,
 \label{form5}
 \ee
 and
 \be
 (V\phi)(\vp_1,\vp_2)=\int \frac{d^3p'_1}{(2\pi)^4i}\frac{d^3p'_2}{(2\pi)^4i}
  I_{inst}\phi(\vp'_1,\vp'_2).
 \label{form5-1}
 \ee
 To obtain $Z(\vp_1,\vp_2)$, we use the definition
 \be
 \om_i\equiv \sqrt{\vp_i^2+m_i^2},\hspace{1cm}i=1,2,3,
 \label{form6}
 \ee
 and rewrite Eq.(\ref{form5}) as follows,
 \bea
 \frac{Z(\vp_1,\vp_2)}{8m_1m_2m_3}&=&\int dx dy \frac{1}{(x^2-\om_1^2+i\epsilon)}
 \frac{1}{(y^2-\om_2^2+i\epsilon)} \frac{1}{\left((P^0-x-y)^2-\om_3^2+i\epsilon\right)}\nn\\
 &=&\frac{1}{4\om_1\om_2\left((P^0-\om_1-\om_2)^2-\om_3^2\right)}+
 \frac{1}{4\om_2\om_3\left((P^0-\om_2+\om_3)^2-\om_1^2\right)}\nn\\
 &+&\frac{1}{4\om_1\om_3\left((P^0-\om_1+\om_3)^2-\om_2^2\right)}+
 \frac{1}{4\om_1\om_3\left((P^0+\om_1+\om_3)^2-\om_2^2\right)}.
 \label{form7}
 \eea
  Simplifying the above result one finds that
 \be
 Z(\vp_1,\vp_2)=\frac{2m_1m_2m_3}{\om_1\om_2\om_3}
 \left(\frac{\om_2+\om_3}{(P^0-\om_1)^2-(\om_2+\om_3)^2}+
 \frac{\om_1+\om_2}{(P^0+\om_3)^2-(\om_1+\om_2)^2}\right).
 \label{formula17}\ee
 Using Eq.(\ref{formula17}) in Eq.(\ref{form4}), we obtain
 \be
 {\cal V}\phi(\vp_1,\vp_2)=\frac{m_1m_2m_3}{\om_1\om_2\om_3}
 \left(\frac{\om_2+\om_3}{\frac{{\cal V}}{2}+\om_2+\om_3}+
 \frac{\om_1+\om_2}{\frac{{\cal V}}{2}+\om_3+\om_1+\om_2}
 \frac{{\cal V}}{{\cal V}+2\om_3}\right)
 (V\phi)(\vp_1,\vp_2),
 \label{Klein}
 \ee
 where ${\cal V}\equiv (P^0-\om_1-\om_2-\om_3)$. Assuming ${\cal V}\ll \om_i$, $i=1,2,3$ we
 have
 \be
 {\cal V}\phi(\vp_1,\vp_2)=\frac{m_1m_2m_3}{\om_1\om_2\om_3}
  (V\phi)(\vp_1,\vp_2),
 \ee
 Therefore, in configuration space the above equation in non-relativistic limit i.e.
 $\om_i \approx m_i$, leads to :
 \be
 H\left(\vx_1,\vx_2,\vx_3\right)\psi\left(\vx_1,\vx_2,\vx_3;t\right)=E
 \psi\left(\vx_1,\vx_2,\vx_3;t\right),
 \ee
 in which $E=P^0-m_1-m_2-m_3$, $\vx_i$ is the position of the $i$-th particle, and
 \be
 H=-\frac{\nabla_{\vx_1}^2}{2m_1}-\frac{\nabla_{\vx_2}^2}{2m_2}-
\frac{\nabla_{\vx_3}^2}{2m_3}+V\left(\vx_1,\vx_2,\vx_3\right),
 \ee
 where
 \be
 V\left(\vx_1,\vx_2,\vx_3\right)=\frac{q_1q_2}{\left|\vx_1-\vx_2\right|}+
 \frac{q_2q_3}{\left|\vx_2-\vx_3\right|}+\frac{q_1q_3}{\left|\vx_1-\vx_3\right|}.
 \ee
 \par
 Now we are ready to examine the BS equation in the
 non-commutative space. Here, the vertices in kernel $I_{ope}$, given in Eq.(\ref{new1}),
 is multiplied by a momentum dependent phase factor similar to
 Eq.(\ref{p2}), e.g. :
 \be
 I_1\to I_1^{NC}=e^{-\frac{i}{2}\theta^2_{\mu\nu} p_2^\mu
 {p'}_2^\nu}e^{-\frac{i}{2}\theta^3_{\mu\nu} p_3^\mu
 {p'}_3^\nu}I_1.
 \ee
 It can be easily verified that the lowest order corrections due to the $\theta$-dependent
 terms are of order $\theta\alpha^4$ and correspond to the instantaneous
 interaction. Since it is assumed that $\theta_{0i}=0$, in the
 above procedure nothing
 changes but the term $(V\phi)(\vp_1,\vp_2)$ becomes
 \be
(V\phi)(\vp_1,\vp_2,\theta)=\int
\frac{d^3p'_1}{(2\pi)^4i}\frac{d^3p'_2}{(2\pi)^4i}
  (I^{NC})_{inst}\phi(\vp'_1,\vp'_2),
  \label{new2}
  \ee
  where
  \be
  (I^{NC})_{inst}=\sum_{i=1}^3(I^{NC}_i)_{inst},
  \label{new3}
  \ee
  and for example
  \be
  (I_1^{NC})_{inst}=\frac{e^{-\frac{i}{2}\theta^2_{ij} \vp_2^i
 {\vp'}_2^j}e^{-\frac{i}{2}\theta^3_{ij} \vp_3^i
 {\vp'}_3^j}}{\left|{\bf K}_{23}\right|^2}.
 \label{inst}\ee
 From Eq.(\ref{new2}), one can easily verify that the Schr\"{o}dinger
 equation in non-commutative space can be modified as follows,
 \be
 i\frac{\partial}{\partial t}\psi\left(\vx_1,\vx_2,\vx;t\right)=
 H\left(\vx_1,\vx_2,\vx\right)\star\psi\left(\vx_1,\vx_2,\vx;t\right),
 \label{new4}
 \ee
 which is the familiar Schr\"{o}dinger equation in NCQM. The lowest order
 $\theta$-dependent term in NCQED corresponds to the
 instantaneous interaction and is of the order $\theta\alpha^4$ (see Eq.(\ref{inst})).
 Therefore, to obtain the lowest order terms in $\theta$ in NCQED,
 it is sufficient to calculate the corresponding terms from Eq.(\ref{new4}) which is
 the non-relativistic limit of the BS equation in non-commutative
 space.
 \par
 Now we define the center of mass coordinates as
 \bea
 \vR&=&\frac{m_1\vx_1+m_2\vx_2+m_3\vx_3}{m_1+m_2+m_3},\nn\\
 \vr_j&=&\vx_j-\vx_3,\hspace{1cm}j=1,2.
 \eea
 Assume the non-commutativity algebra,
 \be
 \left[\vx_j,\vx_j\right]=i\delta_{ij}c_j\theta,
 \ee
 where for simplicity we have assumed that $\theta^i=c_i\theta$.
 This algebra leads to the following algebra for the center of mass coordinates:
 \be
 \begin{array}{l}
 \left[\vr_1,\vr_2\right]=ic_3\theta,\\
 \left[\vr_j,\vR\right]=i\eta_j\theta,\\
 \left[\vr_j,\vr_j\right]=i\kappa_j\theta,\\
 \left[\vR,\vR\right]=i\beta\theta.
\end{array}\hspace{1cm}j=1,2,
 \label{b1}
 \ee
 where,
 \be
 \begin{array}{l}
 \eta_j:=\frac{m_jc_j-m_3c_3}{m_1+m_2+m_3},\\
 \kappa_j:=i(c_j+c_3),\\
 \beta:=\frac{m_1^2c_1+m_2^2c_2+m_3^2c_3}{(m_1+m_2+m_3)^2},
 \end{array}\hspace{1cm}j=1,2.
 \ee
 Considering Eq.(\ref{b1}) and the translation
  \be
 \vr_j\to\vr_j+\frac{1}{2}\eta_j\theta.\vk,
 \ee
 in which $\vk$ is the wave vector of the center of mass \cite{Ho}, one can
 show that
 \bea
 V\star\psi&=&q_1q_3V\left(\vr_1-\frac{i}{2}\kappa_1\theta.\nabla_1-
 \frac{i}{2}c_3\theta.\nabla_2\right)
 +1\leftrightarrow 2 \nn\\
 &+& q_1q_2 V\left(\vr_1-\vr_2-
 \frac{i}{2}\theta.\left(c_1\nabla_1-c_2\nabla_2\right)\right)\psi,
 \label{a1}
 \eea
 where $V(\vx)=\left|\vx\right|^{-1}$.
 Expanding the right hand side of Eq.(\ref{a1}) in terms of $\theta$, one finds
 that Eq.(\ref{new4}) can be written as
 \be
 E\psi=\left(H+H^{NC}_{I}\right)\psi,
 \ee
 where, the operator $H$ is the Hamiltonian of the three body bound state in the
 commutative space and
 \be
 H^{NC}_I=-\frac{q_1q_3}{2}\vt.\frac{\vr_1\times\left(\kappa_1\vp_1+c_3\vp_2\right)}
 {\left|\vr_1\right|^3}+1\leftrightarrow 2+\frac{q_1q_2}{2}\vt.
 \frac{\left(\vr_1-\vr_2\right)\times\left(c_1\vp_1-c_2\vp_2\right)}
 {\left|\vr_1-\vr_2\right|^3}.
 \label{a2}
 \ee
 To obtain Eq.(\ref{a2}) we have used the relation
 \be
 {\bf u}.\theta.{\bf v}=\vt.\left({\bf u}\times {\bf v}\right),
 \ee
 for arbitrary vectors ${\bf u}$ and ${\bf v}$ in
 which $\vt$ is a vector valued parameter defined as follows
 \cite{HaLo}:
 \be
 \vt:=\left(\theta_{23},\theta_{31},\theta_{12}\right).
 \ee
 For example for the bound state of two electrons
 ($c_i=1$, $i=1,2$), and a nucleus ($i=3$), e.g.  the Helium atom,
 $H_I^{NC}$ takes the following form,
 \be
 H^{NC}_I=\frac{Ze^2}{2}\kappa\vt.\left(\frac{{\bf\ell}_1}
 {\left|\vr_1\right|^3}+\frac{{\bf\ell}_2}
 {\left|\vr_2\right|^3}\right)+\frac{e^2}{2}
 \frac{\vt.{\bf\ell}}{\left|\vr\right|^3}+\frac{Ze^2}{2}c_3\vt.\left(\frac{\vr_1\times\vp_2}
 {\left|\vr_1\right|^3}+\frac{\vr_2\times\vp_1}
 {\left|\vr_2\right|^3}\right),
 \label{new5}
 \ee
 where $\vr=\vr_1-\vr_2$, ${\bf\ell}=\vr\times(\vp_1-\vp_2)$ and
 ${\bf\ell}_i=\vr_i\times\vp_i$, $i=1,2$.
   One should note that these terms result in energy shifts of order
$\alpha^4$ (tree level corrections) and they should be added to
the terms with the same order of $\alpha$ in $H$. For example, the
second term of Eq.(\ref{new5}) changes the spin-other-orbit
interaction term \cite{Sh} to
  \be
  u^{nc}_{s-o-o}=-\frac{\alpha}{2m^2}
  \left[\frac{\left(\vs_1+m^2\vt\right).\left(\vr\times\vp_2\right)}{\left|\vr\right|^3}
  +1\leftrightarrow 2\right].
  \ee
 It is interesting to note that if one assign a non-commutativity parameter with
opposite sign relative to that of electron to the nucleus, i.e.
$c_3=-1$ \cite{Ho}, then only the first term in Eq.(\ref{new5})
vanishes and that equation reduces to
 \be
 H^{NC}_I=\frac{e^2}{2}
 \frac{\vt.{\bf\ell}}{\left|\vr\right|^3}+\frac{Ze^2}{2}c_3\vt.\left(\frac{\vr_1\times\vp_2}
 {\left|\vr_1\right|^3}+\frac{\vr_2\times\vp_1}
 {\left|\vr_2\right|^3}\right).
 \label{new6}
 \ee
 Consequently, also in this case, there exist three level corrections for Helium
 atom.
  \par
  To calculate the energy shift resulted by $H^{NC}_I$ given in
  Eq.(\ref{new5}), we consider the state $(1s)(np)$ and assume
  the electron labeled $i=1$ as inner electron and that labeled
  $i=2$ as outer one. Since the inner electron is in the ground
  state and the outer electron in an excited state, we
  can take an unsymmetrized wave function as
  $u_0(\vr_1)u_n(\vr_2)$ in which $u_i$, are approximated by Hydrogen-like
  wave functions. It is known that the error committed in leaving out
  symmetrization is of the order of the ratio of the radii of the
  orbits \cite{Bethe}. Consequently the third term (and similar expressions in the second term)
  in Eq.(\ref{new5}) has no contribution in the energy shift as a
  consequence of their odd parity. Therefore,
  \bea
  \delta E&=&\left<0n\right|H_I^{NC}\left|0n\right>\nn\\
  &\approx&\frac{e^2}{2}(\kappa Z-1)\vt.\left<0n\right|\frac{{\bf
  \ell}_2}{\left|\vr_2\right|^3}\left|0n\right>\nn\\
 &=&f(n,l)\frac{\left|\vt\right|\alpha^4}{\lambda_e^3}m,\hspace{1cm}
 m=-\ell_2,\cdots,\ell_2,
  \label{end1}\eea
  where
  \be
 f(n,l)=\frac{(\kappa
 Z-1)(Z-1)^3}{n^3(2\ell_2+1)(\ell_2+1)\ell_2},
 \ee
 and $\lambda_e$ is the Compton wave length of the electron.
  In the last equality of Eq.(\ref{end1}) we have assumed $\vt=\left|\vt\right|\hat{z}_2$.
   The above calculations are reasonable at least if the outer electron
  is in a highly excited state. Of course these terms can be
  calculated more accurately
  using the method of reference \cite{Breit}.
  Eq.(\ref{end1}) gives a Zeeman effect caused by
  non-commutativity. Since
  the reported uncertainties on the experimental values of atomic levels
$2P$ in Helium atom are about $1$ kHz, from Eq.(\ref{end1}) one
can see that such an uncertainty gives an upper bound ${\frac
{\left|\vt\right|}{{\lambda}_e^2}}\sim 10^{-9}$ or
${\left|\vt\right|}\sim 10^{-33}\mbox{m}^2$.
\section*{Summary}
Although in first sight it seems that Helium atom is not suitable
to be considered as a test on non-commutativity but we found that
the upper bound on $\left|\vt\right|$ obtained in such a system
is even more accurate than that expected for Hydrogen atom and
positronium. For example it is shown that there is not any
correction at the order $\alpha^4$ for the hyper fine splitting
of positronium and the correction to the energy shift are of order
$\frac{\left|\vt\right|}{\lambda_e^3}\alpha^6$ which gives an
upper bound $\frac{\left|\vt\right|}{\lambda_e^2}\sim
\frac{10^{-9}}{\alpha^2}\sim 10^{-5}$ for 1 kHz uncertainty in
experiment \cite{Zeb}. The atomic transition $(2^3P_1)\to
(2^3P_2)$ in Helium atom is measured with an accuracy 1kHz
\cite{Storry}. Eq.(\ref{end1}) shows that the states $(2^3P_1)$
and $(2^3P_2)$ are split themselves and the energy intervals are
of order $\frac{\left|\vt\right|}{\lambda_e^3}\alpha^4$. Thus the
upper bound on $\left|\vt\right|$ obtained here is
$\frac{\left|\vt\right|}{\lambda_e^2}\sim 10^{-9}$.
\section*{Acknowledgement}
The authors would like to thank Dr. S. M. Zebarjad for useful
discussions. The financial support of IUT research council and
IPM are acknowledged.

\end{document}